\documentclass[twocolumn, aps, prb]{revtex4}
\usepackage{graphicx}
\usepackage{amssymb}
\usepackage{amsmath}

\def\lsim{\lower -0.3ex \hbox{$<$} \kern -0.75em \lower 0.7ex \hbox{$\sim$}}
\def\gsim{\lower -0.3ex \hbox{$>$} \kern -0.75em \lower 0.7ex \hbox{$\sim$}}

\newcommand{\GVec}[1]{\mbox{\boldmath$#1$}}

\def\Vec#1{{\bf #1}}
\def\GVec#1{\mbox{\boldmath $#1$}}

\def\vare{\varepsilon}

\begin{document}

\title{Magneto-optical properties of multilayer graphenes}
\author{Mikito Koshino and Tsuneya Ando}
\affiliation{
Department of Physics, Tokyo Institute of Technology\\
2-12-1 Ookayama, Meguro-ku, Tokyo 152-8551, Japan}
\date{\today}

\begin{abstract}
The magneto-optical absorption properties
of graphene multilayers are theoretically studied.
It is shown that the spectrum can be decomposed 
into sub-components effectively identical to the monolayer 
or bilayer graphene, allowing us to understand
the spectrum systematically
as a function of the layer number.
Odd-layered graphenes always exhibit
absorption peaks which shifts in proportion to $\sqrt{B}$,
with $B$ being the magnetic field,
due to the existence of an effective monolayer-like subband.
We propose a possibility of observing
the monolayer-like spectrum even in a mixture of 
multilayer graphene films with various layers numbers.
\end{abstract}

\maketitle

\section{Introduction}

The unusual electronic property of 
the atomically thin graphene films 
has been of great interest.
Recently the optical absorption spectra were measured
in graphene-related systems under magnetic fields.
\cite{Sado06,Zian07,Deac07,Ploc07,Orli07,Henr07}
In this paper we theoretically study magneto-optical spectra
of the graphene multilayer.

The monolayer graphene is a zero-gap semiconductor
with the linear dispersion analogous to the
zero-mass relativistic particle.
In presence of a magnetic field $B$, 
it gives an unusual sequence
of the Landau levels with spacing proportional to $\sqrt{B}$
in both of the electron and hole sides.
\cite{McCl56}
The transport properties in such a unique band structure
were studied and found to be significantly 
different from the conventional system.
\cite{Shon_and_Ando_1998a,Gonzalez_et_al_2001a,Zheng_and_Ando_2002a,Ando_et_al_2002a,Suzuura_and_Ando_2002b,Gusy05}
Recent experimental realizations of monocrystalline graphene
opened the way for direct probing of those unusual properties.
\cite{Novo04,Novo05,Zhan05-2}
The Landau level structure of the single-atomic sheet of graphene
was investigated through the quantum Hall effect \cite{Novo05,Zhan05-2}
and the cyclotron resonance.
\cite{Zian07,Deac07}
The optical response in graphene monolayer 
was theoretically studied.
\cite{Ando_et_al_2002a,Gusy06a,Gusy06b}

The multilayer systems containing few layers of graphene
are also fabricated, \cite{Novo05,Ohta} 
have attracted broad interest as well. \cite{Kope07}
There the interlayer coupling drastically changes 
the structure around the band touching point. 
\cite{Ohta, McCa06, Koshino_and_Ando_2006a, 
Nils06, Guin06, Lati06, Part06, Lu06,Koshino_and_Ando_2007b}
On the other hand, recent observations of the magneto-absorption spectra
of thin epitaxial graphite \cite{Sado06} show 
$\sqrt{B}$-dependent transition peaks just as 
in monolayer graphene.\cite{Zian07,Deac07}
Similar evidences for the linear dispersion
were also found in thicker graphite systems. 
\cite{Ploc07,Orli07,Toy77,Zhou06}
Quite recently the cyclotron resonance was measured in 
graphene bilayer. \cite{Henr07}
In theories, the electronic structure in magnetic fields
has been extensively studied for three-dimensional (3D) 
graphite \cite{McCl56,McCl60,Inou62,Gupt72,Dres74,Naka76}
and for few-layers graphenes.
\cite{McCa06,Guin06,Pere07}
The optical absorption was theoretically
investigated for the bilayer graphene.
\cite{Aber07,Pere07}

Here we study the optical absorption properties
of the AB-stacked multilayer graphenes in magnetic fields
systematically as a function of the layer number.
We decompose the Hamiltonian into 
subsystems effectively identical to
monolayer or bilayer graphene,
\cite{Koshino_and_Ando_2007b}
and express the spectrum as a summation over each of them.
We present in Sec.\ \ref{sec_form}
the Hamiltonian decomposition and the Landau-level structure
of the multilayer graphene
as well as the formulation of the optical absorption.
We show the numerical results in Sec.\ \ref{sec_resu} 
and discussion in Sec.\ \ref{sec_disc}.

\section{Formulation}
\label{sec_form}
We consider a multilayer graphene
composed of $N$ layers of a carbon hexagonal network,
which are arranged in the AB (Bernal) stacking.
The system can be described by a {\bf k}$\cdot${\bf p} Hamiltonian
based on 3D
graphite model.\cite{Wall47,Slon58,McCl57}
The effective models were derived for the monolayer graphene,
\cite{McCl56,DiVincenzo_and_Mele_1984a,Ando_2005a,Ajik93}
the bilayer, \cite{McCa06} and the trilayer and more.
\cite{Guin06,Koshino_and_Ando_2007b}
For the simplicity we include 
the nearest-neighbor intra-layer coupling parameter $\gamma_0$,
and the inter-layer coupling $\gamma_1$ between
A and B atoms located vertically with respect to the layer plane.
The band parameters were experimentally estimated
in bulk graphite as $\gamma_0 \approx 3.16$ eV \cite{Toy77} 
and $\gamma_1 \approx 0.39$ eV. \cite{Misu79} 
The effects of other band parameters neglected here
will be discussed in Sec.\ \ref{sec_disc}.

The low energy spectrum is given by the states 
in the vicinity of $K$ and $K'$ points in the Brillouin zone.
Let $|A_j\rangle$ and $|B_j\rangle$ be the Bloch functions at the $K$
point, corresponding to the $A$ and $B$ sublattices, respectively, of
layer $j$.
For convenience we divide carbon atoms into two groups as
\begin{eqnarray}
 {\rm Group\,\,I :} && B_1, \, A_2, \, B_3, \, \cdots \\
 {\rm Group\,\,II :} && A_1, \, B_2, \, A_3, \, \cdots
\end{eqnarray}
The atoms of group I are arranged along vertical columns
normal to the layer plane, 
while those in group II are above or below
the center of hexagons in the neighboring layers.
The lattice constant within a layer
is given by $a=0.246$ nm and the distance between adjacent layers
$c_0/2 = 0.334$ nm.

If the basis is taken as $|A_1\rangle,|B_1\rangle$;
$|A_2\rangle,|B_2\rangle$; $\cdots$; $|A_N\rangle,|B_N\rangle$, the
Hamiltonian for the multilayer graphene
around the $K$ point becomes 
\begin{eqnarray}
 {\cal H} =
\begin{pmatrix}
 H_0 & V & & & \\
 V^{\dagger} & H_0 & V^{\dagger}& & \\
  & V & H_0 & V & \\
  &  & \ddots & \ddots & \ddots
\end{pmatrix},
\label{eq_H}
\end{eqnarray}
with
\begin{eqnarray}
 H_0 = 
\begin{pmatrix}
 0 & v \pi_- \\ v \pi_+ & 0
\end{pmatrix}, \quad
 V = 
\begin{pmatrix}
 0 & 0 \\ \gamma_1 & 0
\end{pmatrix}.
\end{eqnarray}
where  $\pi_\pm = \pi_x \pm i \pi_y$ with
$\GVec{\pi} = -i\hbar \nabla + e \Vec{A}$, the vector potential $\Vec{A}$, and $v$ is the band velocity of monolayer graphene, 
which is related to the band parameter via
$v = \sqrt{3} a \gamma_0/2\hbar$.
The effective Hamiltonian for $K'$ 
is obtained by exchanging
 $\pi_+$ and $\pi_-$.


%
The Hamiltonian (\ref{eq_H}) can be decomposed into smaller subsystems
for the basis appropriately chosen.\cite{Koshino_and_Ando_2007b}
First, we define the orthonormal sets
%
\begin{eqnarray}
 |\phi_{l}^{\rm (I)}\rangle \!\! &=& \!\!
\psi_l(1) |B_1\rangle + \psi_l(2)|A_2\rangle + 
\psi_l(3)| B_3\rangle + \cdots \! , \quad \nonumber\\
\noalign{\vskip-0.150cm}
\\
\noalign{\vskip-0.250cm}
 |\phi_{l}^{\rm (II)}\rangle \!\! &=& \!\! 
\psi_l(1) |A_1\rangle + \psi_l(2)|B_2\rangle + 
\psi_l(3)| A_3\rangle + \cdots \! , \qquad \nonumber
\end{eqnarray}
%
where
%
\begin{equation}
 \psi_l(j) = \sqrt{\frac{2}{N+1}} \sin j \kappa_l,
\quad
\kappa_l = \frac{\pi}{2}-\frac{l\pi}{2(N+1)} ,
\label{eq_psi}
\end{equation}
%
with
%
\begin{equation}
l = -(N-1), \, -(N-3), \, \dots, \, N-1 .
\end{equation}
%
Here, $l$ is an odd integer when the layer number $N$ is even, while $l$ is even when $N$ is odd, and therefore $l=0$ is allowed only for odd $N$.
\par
%
Next, for $m>0$, we take the basis
%
\begin{eqnarray}
\left\{
(|\phi_{m}^{\rm (II)}\rangle+|\phi_{-m}^{\rm (II)}\rangle)/\sqrt{2}, \quad
(|\phi_{m}^{\rm (I)}\rangle+|\phi_{-m}^{\rm (I)}\rangle)/\sqrt{2},
\right.
\nonumber \\
\left.
(|\phi_{m}^{\rm (I)}\rangle-|\phi_{-m}^{\rm (I)}\rangle)/\sqrt{2}, \quad 
(|\phi_{m}^{\rm (II)}\rangle-|\phi_{-m}^{\rm (II)}\rangle)/\sqrt{2} 
\right\} .
\label{eq_base}
\end{eqnarray}
%
For $m=0$, we take the basis $\{|\phi_0^{\rm (II)}\rangle,\,|\phi_{0}^{\rm (I)}\rangle\}$.
Then, the Hamiltonian has no off-diagonal elements between different $m$'s.
For $m>0$, the sub-Hamiltonian within the basis Eq. (\ref{eq_base}) becomes
%
\begin{eqnarray}
 {\cal H}_{m} = 
\begin{pmatrix}
0 & v \pi_- & 0 & 0 \\
v \pi_+ & 0 & \lambda_m \gamma_1 & 0 \\
0 &  \lambda_m \gamma_1 & 0 & v \pi_- \\
0 & 0 & v \pi_+ & 0 
\end{pmatrix},
\label{eq_Hm}
\end{eqnarray}
%
with
%
\begin{eqnarray}
\lambda_m = 2 \cos \kappa_m ,
\label{eq_lambda}
\end{eqnarray}
%
which is equivalent to the Hamiltonian of bilayer graphene, while the inter-layer coupling $\gamma_1$ is multiplied by $\lambda_m$.
For $m=0$, we have
%
\begin{eqnarray}
 {\cal H}_{m=0} =
\begin{pmatrix}
0 & v \pi_- \\
v \pi_+ & 0 
\end{pmatrix}
,
\label{eq_H0}
\end{eqnarray}
%
which is identical to the Hamiltonian of the monolayer graphene.
These subsystems are labeled as
%
\begin{equation}
\begin{array}{cc}
 m &= 0, 2, 4, \cdots, N-1  \quad {\rm(odd} \,\, N ) , \\
 m &= 1, 3, 5, \cdots, N-1  \quad {\rm(even} \,\, N ) . 
\end{array}
\end{equation}
%
\par
%
The eigenstate of a finite-layered graphene
can be regarded as a part of a standing wave in 3D limit, 
which is a superposition
of opposite traveling waves with $\pm k_z$.
The quantity $\kappa$ $(=\kappa_m)$ in our representation 
corresponds to the 3D wave number via $\kappa = |k_z|c_0/2$.
Thus the monolayer-type subband $\kappa=\pi/2$ is related to a $H$ point in the 3D Brillouin zone,
while no states exactly correspond to $k_z=0$
since $\kappa$ never becomes zero.


The Landau levels of the monolayer-type states are given by
%
\begin{equation}
\varepsilon_{sn} = s \Delta_B \sqrt{n} ,
\end{equation}
%
with $n=0,1,\dots$ and $s=\pm$, where $s =+$ and $-$ represent the electron and hole bands, respectively, and only $s=+$ is allowed for $n=0$.
\cite{McCl56}
Here $\Delta_B$ is the magnetic energy, 
defined by
\begin{equation}
 \Delta_B = \sqrt{2\hbar v^2 eB}.
\end{equation}

The Landau-level structure of the bilayer-type Hamiltonian (\ref{eq_Hm}) 
was obtained previously\cite{Ando_2007b} and 
can be analytically derived 
by noting that $\pi_\pm$ are associated with the ascending / descending
operators of the Landau levels \cite{Guin06}
in a similar way to that for 3D graphite.
\cite{McCl60,Inou62}
The eigenfunction can be written as
\begin{equation}
 (c_1 \varphi_{n-1,k},\, c_2 \varphi_{n,k},\, c_3 \varphi_{n,k},\, 
c_4 \varphi_{n+1,k}) ,
\label{eq_wave}
\end{equation}
with $n\ge-1$ and amplitudes $c_i$.
Here $\varphi_{n,k}(x,y)$
is the wavefunction of the $n$th Landau 
level in conventional two-dimensional system,
given in the Landau gauge $\Vec{A} = (0, Bx)$ by
$\varphi_{n,k}= i^n (2^n n!\sqrt{\pi}l)^{-1/2}e^{iky} e^{-z^2/2}H_n(z)$ 
with $z = (x+kl^2)/l$ and $H_n$ being the Hermite polynomial.
We define $\varphi_{n,k} \equiv 0$ for $n < 0$.

For $n\ge 1$, 
the Hamiltonian matrix for the vector $(c_1,c_2,c_3,c_4)$
then becomes
\begin{eqnarray}
\begin{pmatrix}
0 & \Delta_B \sqrt{n} & 0 & 0 \\
\Delta_B \sqrt{n} & 0 & \lambda \gamma_1 & 0 \\
0 &  \lambda \gamma_1 & 0 & \Delta_B \sqrt{n+1} \\
0 & 0 & \Delta_B \sqrt{n+1} & 0 
\end{pmatrix},
\label{eq_4x4}
\end{eqnarray}
where the index of $\lambda_m$ is dropped.
This immediately gives four eigen values
\begin{eqnarray}
\vare_{n,\mu,s} &=&  
\frac{s}{\sqrt{2}} \bigg[
(\lambda\gamma_1)^2 + (2n+1)\Delta_B^2  \nonumber\\
&& \hspace{-10mm}
 + \mu \sqrt{(\lambda\gamma_1)^4 
+ 2(2n+1)(\lambda\gamma_1)^2\Delta_B^2+\Delta_B^4}
\bigg]^{1/2}, \quad
\label{eq_bilayer_LL}
\end{eqnarray}
where $\mu = \pm$ correspond to the higher and lower subbands
in the limit of zero magnetic field, respectively. \cite{McCa06}
In the following we use the notation $\mu = H,L$ instead of $+,-$ to avoid the confusion with $s=\pm$. 
The eigen states can be labeled by $n$, $\mu$, $s$, and $k$.

For $n=0$, the first component of the wave function
(\ref{eq_wave}) disappears
and we have only three levels,
\begin{eqnarray}
\vare_{0,L} &=& 0 , \\
\vare_{0,H,s} &=& s \sqrt{\gamma_1^2 + \Delta_B^2}.
\end{eqnarray}
At $n=-1$ only the last component survives in (\ref{eq_wave})
so that we have only a single level in the lower subband, 
$\vare_{-1,L} = 0$ (the level $\vare_{-1,H}$ does not exist).

In small magnetic fields, 
the Landau levels for the lower subband
in the region $\vare \ll \lambda\gamma_1$
are approximately given by $\vare_{n,L,s} \approx s (\hbar eB /m^*) \sqrt{n(n+1)}$
with the effective mass $m^* = \lambda\gamma_1/(2v^2)$. \cite{McCa06}
Thus the level spacing shrinks much faster in $B \rightarrow 0$
than that in the monolayer $\propto\sqrt{B}$.
The ratio of the first gap of the bilayer-type subband, $\hbar eB /m^*$,
to that of the monolayer, $\Delta_B$,
is given by $\Delta_B /(\lambda\gamma_1)$.

\begin{figure}
\begin{center}
 \leavevmode\includegraphics[width=80mm]{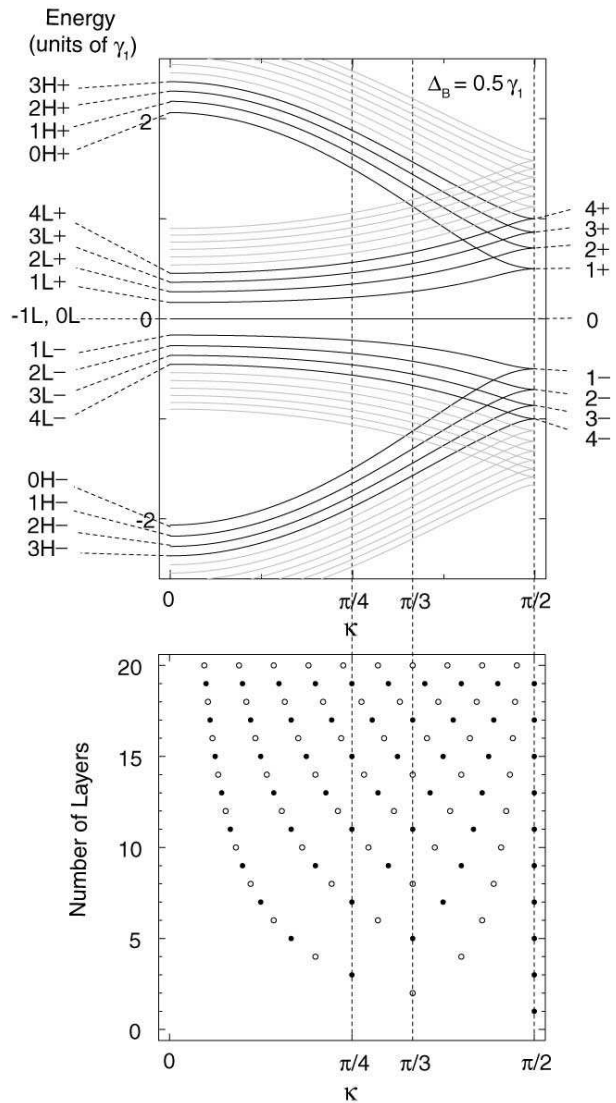}
\end{center}
\caption{(top) Landau levels of the bilayer-type subband
as a function of $\kappa$ with $\lambda = 2\cos \kappa$.
Magnetic field strength is taken as $\Delta_B/ \gamma_1 = 0.5$.
(bottom) Lists of $\kappa$ in $N$-layered graphene.
Empty and filled circles represent even and odd $N$'s, respectively.
}
\label{fig_LL}
\end{figure}

Figure \ref{fig_LL} shows the Landau levels 
of the bilayer-type Hamiltonian 
as a function of $\kappa$, in the magnetic field
given by $\Delta_B/\gamma_1 = 0.5$.
The bilayer levels become those of two independent monolayers at $\kappa=\pi/2$, where the effective inter-layer coupling $\lambda\gamma_1$ vanishes.
The levels become flat around $\kappa = 0$,
where $d\lambda/d\kappa$ vanishes.
In the bottom panel we show the list of $\kappa$
for every layer number $N$.
The top and bottom panels share the horizontal axis;
the Landau levels in the specific point in the bottom panel
are shown directly above.

The velocity operator for the sub-Hamiltonian ${\cal H}_m$
is given by
$v_x = -(i/\hbar)[x,{\cal H}_m] = \partial {\cal H}_m/ \partial \pi_x$.
There are no matrix elements connecting different $m$'s.
For bilayer-type subband,
$v_x$ has a non-zero matrix element
only between the Landau levels with $n$
and $n\pm 1$ for arbitrary combinations of $\mu=H, \, L$
and $s=\pm$.
This is explicitly written as
\begin{eqnarray}
 \langle n'\!,\mu'\!,s' ; k' | v_x | n,\mu,s ; k\rangle
& \!\! = \!\! & 
v \delta_{k,k'}
\left[
(c_1'^* c_2 + c_3'^* c_4)\delta_{n,n'-1} \right.
\nonumber\\
& \!\!\!\! &
\left. 
+ (c_2'^* c_1 + c_4'^* c_3)\delta_{n,n'+1}
\right] , \qquad
\end{eqnarray}
where $c_i$ and $c_i'$ are the eigenvectors of the matrix (\ref{eq_4x4}),
corresponding to the Landau levels $(n,\mu,s)$ and $(n',\mu',s')$,
respectively.
For the monolayer-type band, we have
\begin{equation}
\langle n'\!,s' ; k' | v_x | n,s ; k\rangle = v \delta_{k,k'}
\left[
c_1'^* c_2 \delta_{n,n'-1} + c_2'^* c_1 \delta_{n,n'+1}
\right],
\end{equation}
where $(c_1,c_2)$ is $(0,1)$ for $n=0$ and $(s,1)/\sqrt{2}$
for $n\ge 1$ for $K$-point. \cite{Shon_and_Ando_1998a}


To estimate the optical absorption intensity,
we calculate the real part of the dynamical conductivity
$\sigma_{xx}(\omega)$.
The relative transmission of the sheet to the vacuum,
for the linearly polarized light 
incident perpendicular to the plane, 
is related to this quantity via \cite{Ando75}
\begin{equation}
 T = \Big| 1+ \frac{2\pi}{c}\sigma_{xx}(\omega) \Big|^{-2}
\approx 1- \frac{4\pi}{c}{\rm Re}\,\sigma_{xx}(\omega).
\label{eq:Transmission}
\end{equation}
As will be shown below, the expansion is valid except in thick multilayer graphenes for which the absorption is significant.
The dynamical conductivity can be written in usual manner as
\begin{eqnarray}
 \sigma_{xx}(\omega) = 
\frac{e^2\hbar}{iS}
\sum_{\alpha,\beta}
\frac{f(\vare_\alpha)-f(\vare_\beta)}{\vare_\alpha- \vare_\beta}
\frac{| \langle\alpha | v_x | \beta\rangle |^2}
{\vare_\alpha- \vare_\beta+\hbar\omega+i\delta},
\label{eq_sigma_xx}
\end{eqnarray}
where $S$ is the area of the system,
$v_x$ is the velocity operator, 
$\delta$ is the positive infinitesimal,
$f(\vare)$ is the Fermi distribution function,
and
$| \alpha\rangle$ and $\vare_\alpha$ describe the eigenstate and
the eigen energy of the system.

In the simplest approximation,
we include the disorder effect by
replacing $\delta$ with
the phenomenological constant $\hbar/\tau \equiv 2\Gamma$
and taking the ideal eigenstates as $\alpha$, $\beta$.
The conductivity can then be written as a summation
over all the contributions of the subsystems,
which are independently calculated.
Correspondingly, we compute the density of states 
per unit area as
\begin{equation}
 D(\vare) = 
-\frac{1}{\pi S}
{\rm Im} 
\sum_\alpha 
\frac{1}{\vare - \vare_\alpha + i\Gamma},
\end{equation}
with the ideal eigenstates $\alpha$.

The dynamical conductivity at zero magnetic field
was calculated for the monolayer \cite{Ando_et_al_2002a,Gusy06a}
and the bilayer graphene. \cite{Aber07} 
For the ideal monolayer at $\vare_F = 0$,
the expression apart from $\omega = 0$
becomes a frequency-independent value \cite{Ando_et_al_2002a,Gusy06a}
\begin{equation}
{\rm Re} \, \sigma_{xx} (\omega)  = \frac{g_vg_s}{16}\frac{e^2}{\hbar},
\label{eq_sigma_mono}
\end{equation}
with $g_v = 2$ being the valley ($K,K'$) degeneracy and $g_s = 2$ the spin degeneracy.
Note that the dynamical conductivity has a singularity at $(\varepsilon_F,\omega)=(0,0)$, which is removed if level-broadening effect is included properly.\cite{Ando_et_al_2002a}
The expression for the effective bilayer Hamiltonian (\ref{eq_Hm})
with $\vare_F=0$ is given by \cite{Aber07} 
\begin{eqnarray}
 {\rm Re} \, \sigma_{xx} (\omega)  = \frac{g_vg_s}{16}\frac{e^2}{\hbar}
\left[
\frac{2\lambda^2\gamma_1^2}{\hbar^2\omega^2}\theta(\hbar\omega -\lambda\gamma_1)
\right.\nonumber\\
\left.
+ \frac{\hbar\omega-2\lambda\gamma_1}{\hbar\omega-\lambda\gamma_1}
\theta(\hbar\omega -2\lambda\gamma_1)
+ \frac{\hbar\omega+2\lambda\gamma_1}{\hbar\omega+\lambda\gamma_1}
\right],
\label{eq_sigma_bi}
\end{eqnarray}
where $\theta(x) = 1$ for $x>0$ and $\theta(x)=0$ for $x<0$.
The above shows that the typical value of the real part of the conductivity for $\hbar\omega/\gamma_1<1$ is $(g_vg_s/16)(e^2/\hbar)$ per layer.
By noting that $e^2/\hbar c\approx1/137$, we see that the expansion in Eq.\ (\ref{eq:Transmission}) is valid roughly for $N\,\lsim\,20$.

\section{Numerical Results}
\label{sec_resu}

\begin{figure}
\begin{center}
 \leavevmode\includegraphics[width=70mm]{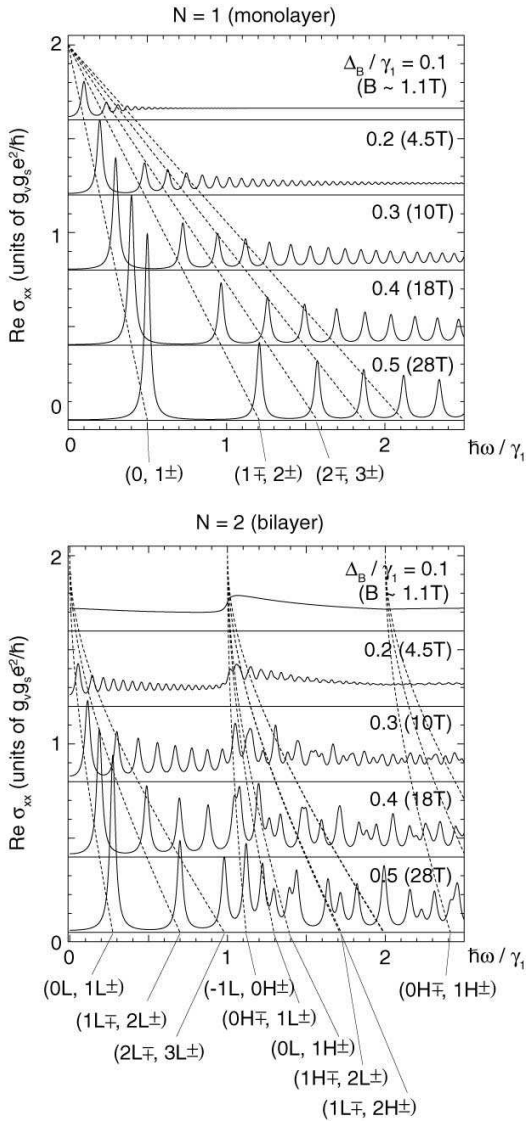}
\end{center}
\caption{
Real part of the dynamical conductivity of 
the monolayer (top) and bilayer (bottom) graphenes
plotted against the frequency $\omega$,
calculated for different magnetic fields (specified by $\Delta_B$)
and $\Gamma/\gamma_1 = 0.01$.
Dashed curves indicate the transition energies
between several Landau levels in the ideal limit.
}
\label{fig_absp1-2}
\end{figure}

Figure \ref{fig_absp1-2} shows the plots of ${\rm Re}\,\sigma_{xx}(\omega)$
for the monolayer and bilayer graphenes
in several magnetic fields.
Here we take $\Gamma/\gamma_1 = 0.01$, $\vare_F = 0$, and
zero temperature. 
Dotted lines penetrating panels 
represent the transition energies between 
several specific Landau levels as a continuous function of $\Delta_B$.
The peak positions of each panel correspond to the
intersections of those and the bottom line of the panel.

In the monolayer the peak position obviously shifts in proportion to
$\sqrt{B}$ (i.e., $\propto \Delta_B$).
In the limit of vanishing magnetic field, the conductivity eventually becomes the value given by Eq.\ (\ref{eq_sigma_mono}).
The spectrum in the bilayer is rather complicated;
starting from $\omega=0$,
we first see the series of 
the transition peaks within $L$ bands from $\omega = 0$,
and then those between $L$ and $H$ enter for $\hbar\omega \,\gsim\, \gamma_1$
and lastly 
those within $H$ bands for $\hbar\omega \,\gsim\, 2 \gamma_1$.
The every peak position behaves as a linear function of $B$ $(\propto \Delta_B^2)$ 
in weak fields
but it switches over to $\sqrt{B}$-dependence as the corresponding energy 
is going out of the parabolic band region.
In small fields the peaks are smeared out more easily in the bilayer 
than in the monolayer.
The conductivity converges to the zero-field curve with a step-like structure at $\vare = \gamma_1$,
which is expressed as Eq.\ (\ref{eq_sigma_bi}) in the clean limit.

\begin{figure}
\begin{center}
 \leavevmode\includegraphics[width=80mm]{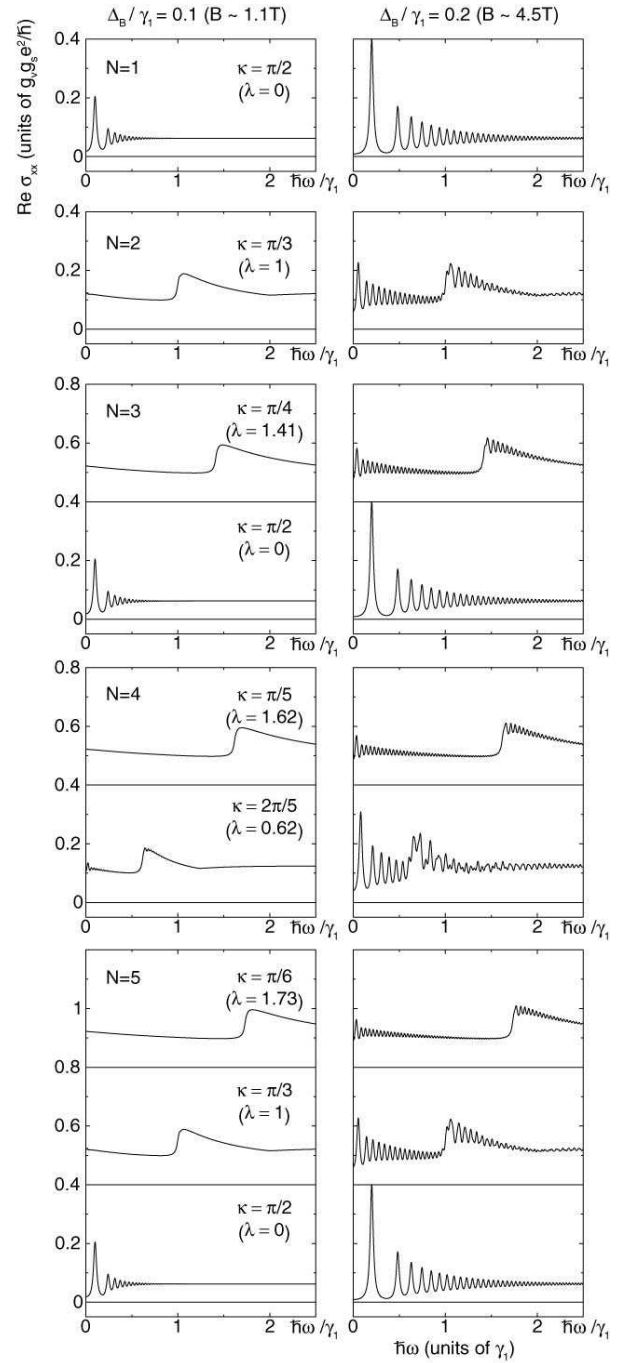}
\end{center}
\caption{
Real part of the dynamical conductivity of $N=1,2, \cdots, 5$-layered
graphenes as functions of the frequency,
calculated for two different magnetic fields
$\Delta_B/\gamma_1 = 0.1$ and $0.2$.
We take $\Gamma/\gamma_1 = 0.01$.
}
\label{fig_absp1-5}
\end{figure}

Figure \ref{fig_absp1-5} shows
the plots of Re$\sigma_{xx}(\omega)$
from $N=1$ to 5
with two different magnetic fields 
with $\Delta_B/\gamma_1 = 0.1$ and $0.2$.
We again take $\Gamma/\gamma_1 = 0.01$, $\vare_F = 0$, and
zero temperature. 
The results are shown separately for each subband.
In every odd layers the monolayer-type subband gives 
the identical spectrum.
All other bilayer-types give different spectra depending on $\lambda$.
The quantized feature is more easily resolved in 
a subband with a smaller $\lambda$, 
because of its narrower level spacings.
In zero field limit, every bilayer-type spectrum has
a step at $\vare \sim \lambda\gamma_1$, where the
excitation between $L$ and $H$ bands starts.

\begin{figure}
\begin{center}
 \leavevmode\includegraphics[width=80mm]{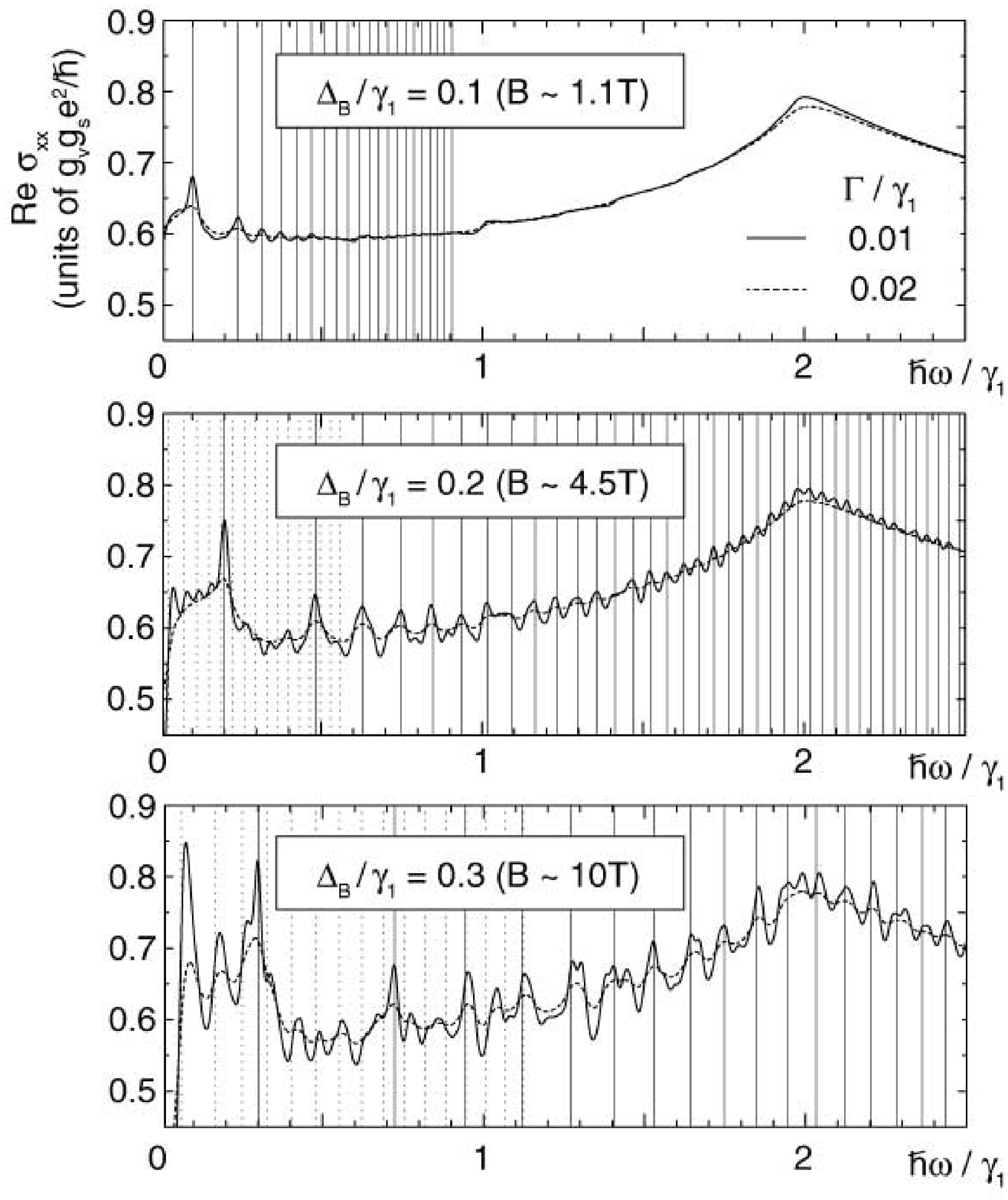}
\end{center}
\caption{
Plots of Re $\sigma_{xx}(\omega)$
averaged over the layer numbers from $N=1$ to 20,
for magnetic fields with $\Delta_B/\gamma_1 = 0.1$, 0.2 and 0.3.
Vertical solid and dashed lines represent
the ideal transition energies for
the monolayer-type ($\kappa=\pi/2$) and the bilayer-type ($\kappa=0$)
subbands, respectively.
}
\label{fig_absp_tot}
\end{figure}

\begin{figure}
\begin{center}
 \leavevmode\includegraphics[width=80mm]{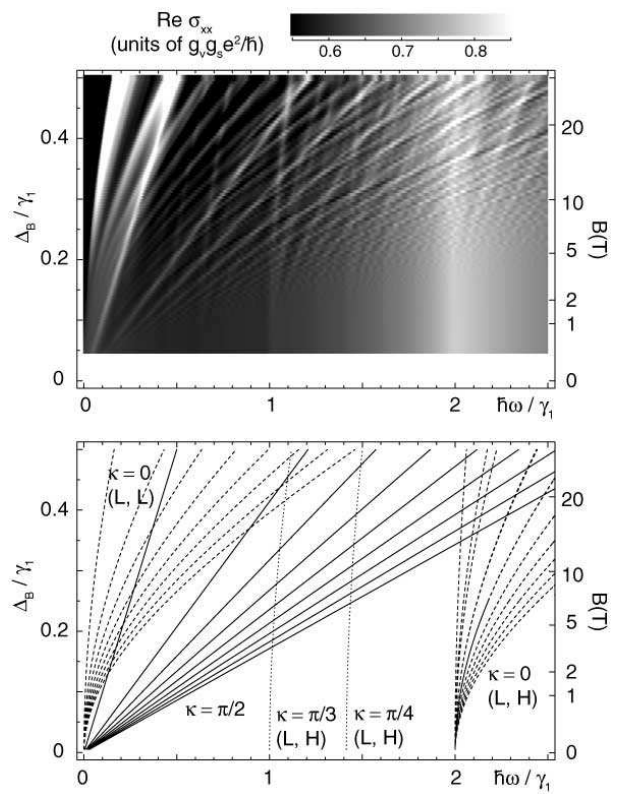}
\end{center}
\caption{
(Top) Density plot of ${\rm Re}\,\sigma_{xx}(\omega,\Delta_B)$
averaged over the layer numbers from $N=1$ to 20.
$\Gamma$ is set to $0.01\gamma_1$.
(Bottom) Corresponding plot for the transition energies
between the several Landau levels.
}
\label{fig_absp_tot_dnst}
\end{figure}

It is intriguing to consider how the absorption spectrum looks like
when the sample is a mixture of thin graphene films
with various layer numbers.
One might think the discreteness of $\kappa$ is easily smeared out
and we just get the 3D limit, but it is not always the case as we will
show in the following.
We here calculate the dynamical conductivity 
averaged over the samples $N=1,2,3,\cdots,20$.
We show in Fig.\ \ref{fig_absp_tot} plots of 
${\rm Re}\,\sigma_{xx}(\omega)$ for different magnetic fields with
$\Delta_B/\gamma_1 = 0.1$, 0.2, and 0.3
and in Fig.\ \ref{fig_absp_tot_dnst} a gray-scale plot
of ${\rm Re}\,\sigma_{xx}(\omega,\Delta_B)$.

Surprisingly we still see the series of peaks in the monolayer graphene
$\hbar\omega \propto \sqrt{B}$.
This comes from the monolayer-type subbands which appears
in every odd layered graphene.
The visibility of the monolayer-type signal depends
on the ratio of the number of monolayer-type subbands to 
the total; in the present case, this is 10 to 110.
We expect that the signal of monolayer gradually becomes invisible
as the maximum layer number $N_{\rm max}$ becomes larger, 
because the total subband number increases as 
$\propto N_{\rm max}^2$
while the number of monolayer-type as $\propto N_{\rm max}$.

We have another set of dominant peaks, which can be identified as
the bilayer-type with $\kappa =0$ $(\lambda=2)$.
Although there is no subband which exactly takes this value,
many subbands around $\kappa \sim 0$
have almost the same peak positions as the Landau level
is flat against $\kappa$ there
and gives similar spectra.
Every peak shifts upward with respect to 
the original position of $\kappa = 0$,
since the Landau level spacing is generally wider for larger $\kappa$.
Unlike the monolayer-type signal,
this would survive even in the 3D limit,
since the finite region in $\kappa$ (not a point) 
can contribute to this spectrum.
When decreasing the magnetic field, however,
the bilayer-type peaks are immediately blurred
due to rapid $B$-linear dependence, while
the monolayer peaks survive even in relatively smaller magnetic field.
In zero-field limit, we are left with a bump at $\hbar\omega = 2\gamma_1$,
which comes from the $H$-$L$ transition step 
of the bilayer-type subbands with $\kappa \sim 0$.

Just in the same way as the monolayer-type subband ($\kappa=\pi/2$) 
appears in every two layers,
the bilayer-type subband with $\kappa=\pi/3$ enters
in every three layers ($N=2,5,8,\cdots$)
and that with $\kappa=\pi/4$ 
in every four layers ($N=3,7,11,\cdots$).
We can see the $H$-$L$ transition peaks 
of those $\kappa$'s 
in Fig. \ref{fig_absp_tot_dnst},
while $L$-$L$ peaks are hidden by other dominant contributions.

\begin{figure}
\begin{center}
 \leavevmode\includegraphics[width=80mm]{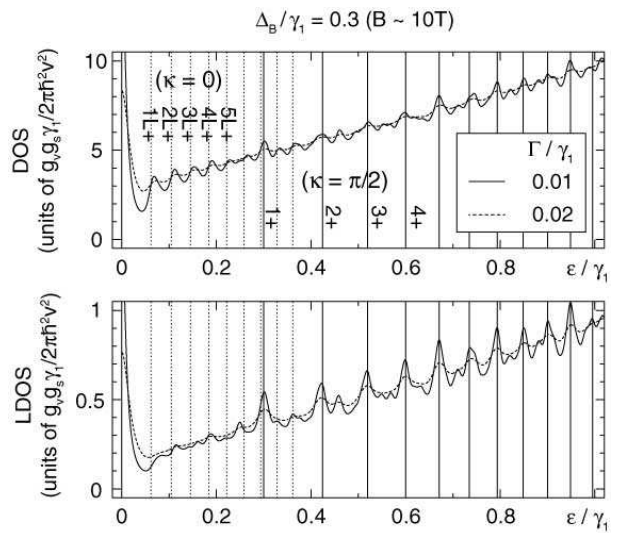}
\end{center}
\caption{
(Top) Density of states averaged over the 
layer numbers $N=1$ to 20, at $\Delta_B/\gamma_1=0.3$.
(Bottom) Corresponding plot for 
the local density of states
on the top layer.
Vertical solid and dashed lines indicate
the ideal Landau level energies for
the monolayer-type ($\kappa=\pi/2$) and the bilayer-type ($\kappa=0$)
subbands, respectively.
}
\label{fig_dos}
\end{figure}

\begin{figure}
\begin{center}
 \leavevmode\includegraphics[width=80mm]{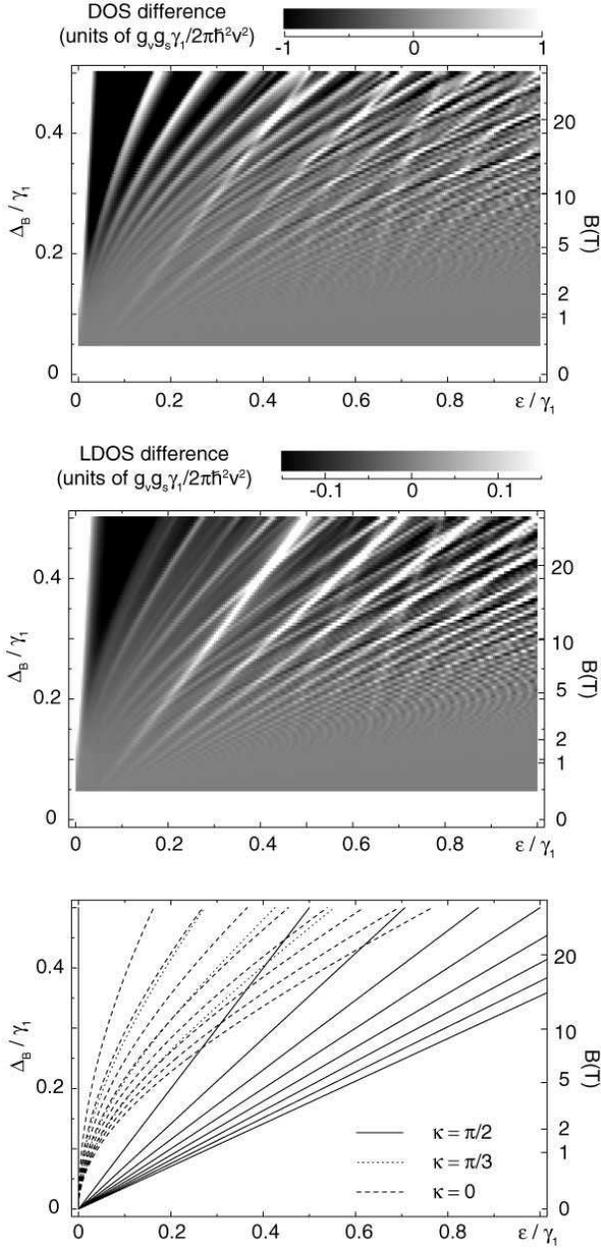}
\end{center}
\caption{
(Top) Density of states (measured from the zero-field value)
averaged over the layer numbers from $N=1$ to 20,
plotted in $(\vare,\Delta_B)$-plane.
$\Gamma$ is set to $0.01\gamma_1$.
(Middle) Similar plot for the local density of states
on the top layer.
(Bottom) Ideal Landau level energies
corresponding to several dominant peaks in above two panels.
}
\label{fig_dos_dnst}
\end{figure}

The similar analysis is available for the density of states (DOS).
In Fig.\ \ref{fig_dos}, the top panel shows DOS averaged 
over $N=1,2,3,\cdots,20$ as a function of the Fermi energy.
The bottom panel shows the corresponding plot 
for the local density of states (LDOS) on the {\it top layer},
defined by the number of states per unit energy width and per unit area 
on the layer.
We also present in Fig.\ \ref{fig_dos_dnst} the two-dimensional plots
of DOS and LDOS on $(\vare,\Delta_B)$-plane,
where the gray-scale shows the relative value from the zero magnetic field.

In DOS, 
we observe the several peaks coming from the monolayer-type subband
similarly to the optical absorption spectra.
The peaks from the bilayer-type subband with $\kappa=0$ 
become prominent in the high-field region $\Delta_B/\gamma_1 \,\gsim\, 0.2$.
In LDOS, interestingly, the peaks of the monolayer-type subband 
are much more pronounced,
while those of $\kappa = 0$ are strongly suppressed.
This can be understood by the wave function defined by 
Eq.\ (\ref{eq_base}). If we look at a state in the subband $m$ 
in $N$-layered graphene, the wave amplitude on the top layer $(j=1)$
always acquires the factor $\psi_m(1) = \sqrt{2/(N+1)}\sin\kappa_m$.
Obviously this takes maximum in the monolayer-type
($\kappa=\pi/2$) and zero at $\kappa=0$,
and thus the monolayer-type state contributes the most to the surface LDOS.
The bilayer-type signals of the subbands $\kappa=\pi/3$ and $\pi/4$
are also visible in LDOS
while they are hidden by $\kappa=0$ in DOS.

\section{Discussion}
\label{sec_disc}

Recently the optical absorption spectrum was measured 
in the epitaxial thin graphite films
and the monolayer-like signal was observed,
while the detail profile of the system remains unclear.
Similar $\sqrt{B}$-dependent features 
were also observed in 
the samples containing high-number of graphene layers 
($\sim$ 100) grown on SiC substrate,\cite{Ploc07}
and in a thin graphite sample of thickness $\sim$ 100 nm
exfoliated from highly-oriented pyrolytic
graphite.\cite{Orli07}
Those results are nontrivial because,
if the system is a real three-dimensional bulk graphite,
the spectrum would be contributed mainly from the states
around $k_z = 0$ ($\kappa=0$ in our discussion) 
where the Landau levels are flat with respect to $k_z$.
One possible scenario for this is that
the system can be regarded as
a compound of multilayer fragments with various small 
layer numbers,
and the monolayer-like spectra of all the odd layers are observed.
It should also be mentioned that the local density of states on the 
surface of graphite was observed in the experiment.
\cite{Mats05,Niim06}
Our calculation predicts that 
the pronounced monolayer-type spectrum would be observable
in a multilayer graphene.


While we adopted a simplified effective-mass
model in which only $\gamma_0$ and $\gamma_1$ are included,
here we briefly mention the effects of other hopping parameters.
The parameter $\gamma_3$ neglected here couples
group II atoms on neighboring layers.
This is responsible for the trigonal warping of the band dispersion, 
but gives only a slight shift in the Landau level energies
except for the low energy region ($<$ 10 meV).
\cite{Gupt72,Dres74,Naka76}
Therefore, it would hardly affect the peak positions in the absorption spectra
while may modify the amplitudes through the matrix element changes.
The parameter $\gamma_4$ couples group 
I and II atoms sitting on the neighboring layers, such as
$A_j \leftrightarrow A_{j+1}$ or $B_j \leftrightarrow B_{j+1}$.
This parameter introduces a small electron-hole asymmetry 
in the band structure, but
does not change the qualitative feature of the low-energy
spectrum.\cite{McCl57}

We also neglected the vertical hopping between the
second-nearest neighboring layers for group II and I atoms,
which are parameterized by
$\gamma_2$ and $\gamma_5$, respectively. 
Including those parameters mainly shifts the zero energy
(the band touching point) upward or downward,
depending on each subsystem.\cite{Lati06,Part06,Lu06}
in the tight-binding model \cite{Part06,Lu06}
and the density functional theory \cite{Lati06}
estimate the shift $\delta E$ at the order of 10 meV.
In 3D limit, this corresponds to the band dispersion along $k_z$-direction. 
\cite{McCl56,McCl60,Inou62,Gupt72,Dres74,Naka76}
The zero-energy shift leads to the electron or hole doping,
and gives a change of the absorption spectrum 
in the region $\hbar\omega \,\lsim\, 2\delta E$.


The effective mass model is
no longer valid when the energy is as high
as the intra-layer coupling $\gamma_0 \sim 3$ eV.
The lattice effect appears as
trigonal warping in the band dispersion
in higher energies, \cite{Ajik96,Sait00,Grun03}
while this should be distinguished from the trigonal warping
discussed above, which is due to the extra band parameter
within the effective mass model.
The frequency region covered in our calculation,
$\hbar\omega \lsim 2.5\gamma_1 \sim 1$ eV,
roughly corresponds to the energy region
$|\vare| \lsim 0.5$ eV.
The deviation in eigen energy is estimated at 5\% at $\vare = 0.5$ eV
and can be treated perturbationally, \cite{Ajik96}
although it grows as the energy increases out of this region. 
This anisotropy constitutes a major part of the chirality dependence of optical spectra in carbon nanotubes, enabling the assignment of the structure of individual nanotubes.\cite{Bachilo_et_al_2002a}


Lastly, while our model is based on the bulk 3D graphite,
it should be noted that
the band parameters in few-layered graphenes
are not exactly the same as those for the bulk graphite,
but generally vary depending on the layer number. \cite{Ohta}
There is a theoretical attempt to obtain 
accurate electronic structures for few-layered graphenes,
using the density functional theory
with the local density approximation. \cite{Lati06}
The calculation beyond the 
local density approximation was also 
proposed, which properly treats nonlocal van der Waals interaction
coupling graphene layers in the density functional framework. 
\cite{Rydb00}
The study of the optical absorption in a refined band model
is left for a future work.


In conclusion, we have presented a systematic  
study of the optical absorption properties and the density of states
in the multilayer graphenes as a function of layer numbers.
The spectrum can be understood through the decomposition
into sub-components, each of which is equivalent to the monolayer graphene
or the bilayer graphene with single parameter $\kappa$.
We proposed that the monolayer-like spectra is possibly observed
in the mixture of the multi-layered graphene,
contributed by the effective monolayer subbands existing in every odd-layered
graphene.

\section*{ACKNOWLEDGMENTS}

The authors acknowledge helpful interactions with 
E. A. Henriksen, Z. Jiang, K. F. Mak, P. Kim, and T. F. Heinz.
This work has been supported in part by the 21st Century COE Program at
Tokyo Tech \lq\lq Nanometer-Scale Quantum Physics'' and by Grants-in-Aid
for Scientific Research and Priority Area 
``Carbon Nanotube Nano-Electronics'' from the Ministry of Education, 
Culture, Sports, Science and Technology, Japan.


\end{document}